\documentclass[12pt,preprint]{aastex}

\def\ltsima{$\; \buildrel < \over \sim \;$}
\def\lsim{\lower.5ex\hbox{\ltsima}}
\def\gtsima{$\; \buildrel > \over \sim \;$}
\def\gsim{\lower.5ex\hbox{\gtsima}}
\def\rs{$R_S = 1.10 \pm 0.07$~$R_\odot$}
\def\rp{$R_P = 1.06 \pm 0.08$~$R_{\rm Jup}$}
\def\ms{$M_S = 1.025^{+0.135}_{-0.130}$~$M_\odot$}
\def\mp{$M_P = 0.57 \pm 0.12$~$M_\odot$}

\def\kms{\ifmmode{\rm km\thinspace s^{-1}}\else km\thinspace s$^{-1}$\fi}

\shortauthors{Holman et al.}
\shorttitle{OGLE-TR-10}

\begin{document}

\bibliographystyle{apj}

\title{The Transit Light Curve Project.\\
IV.~Five Transits of the Exoplanet OGLE-TR-10b
}

\author{
  Matthew~J.~Holman\altaffilmark{1},
  Joshua~N.~Winn\altaffilmark{2}, 
  Cesar~I.~Fuentes\altaffilmark{1}, 
  Joel~D.~Hartman\altaffilmark{1}, 
  K.~Z.~Stanek\altaffilmark{3},
  Guillermo~Torres\altaffilmark{1},
  Dimitar~D.~Sasselov\altaffilmark{1},
  B.~Scott~Gaudi\altaffilmark{1,3},
  R.~Lynne~Jones\altaffilmark{4},
  Wesley~Fraser\altaffilmark{5}
}

\altaffiltext{1}{Harvard-Smithsonian Center for Astrophysics, 60
  Garden Street, Cambridge, MA 02138; mholman@cfa.harvard.edu}

\altaffiltext{2}{Department of Physics, and Kavli Institute for
  Astrophysics and Space Research, Massachusetts Institute of
  Technology, Cambridge, MA 02139}

\altaffiltext{3}{Department of Astronomy, The Ohio State University,
  Columbus, OH 43210}

\altaffiltext{4}{National Research Council of
Canada, Hertzberg Institute for Astrophysics, Victoria, BC, V9E VE7, Canada}

\altaffiltext{5}{Department of Physics and Astronomy, University of
Victoria, Victoria, BC V6T~1Z1, Canada}

\begin{abstract}
We present $I$ and $B$ photometry of five distinct transits of the
exoplanet OGLE-TR-10b. By modeling the light curves, we find the
planetary radius to be \rp~and the stellar radius to be \rs. The
uncertainties are dominated by statistical errors in the photometry.
Our estimate of the planetary radius is smaller than previous
estimates that were based on lower-precision photometry, and hence the
planet is not as anomalously large as was previously thought. We provide
updated determinations of all the system parameters, including the
transit ephemerides.
\end{abstract}


\section{Introduction}

Apart from Mercury and Venus, we know of 10 planets that transit their
parent stars as viewed from Earth.  Three of the transiting extrasolar
planets (HD~209458b, HD~149026b, and HD~189733b) were discovered by
observing radial velocity variations of the parent star and then
searching for the photometric signal of transits
\citep{Mazeh.2000,Charbonneau.2000,Henry.2000,Sato.2005,Charbonneau.2005}. In
the other cases, the photometric signals were discovered first, and
then confirmed as planetary transits through radial velocity studies
\citep{Udalski.2002a,Udalski.2002b,Udalski.2002c,Udalski.2003,Udalski.2004,
Bouchy.2004,Bouchy.2005a,Konacki.2003a,Konacki.2003b,Konacki.2004,Konacki.2005,
Alonso.2004,Pont.2004,McCullough.2006}.
Regardless of the order of events, the combination of photometry and
dynamical measurements allows the mass and radius of the planet (and
hence its mean density) to be determined.

These measurements set the stage for a host of more subtle
measurements of effects such as planetary atmospheric absorption
lines, thermal emission, spin-orbit alignment, and timing anomalies,
as reviewed recently by \citet{Charbonneau.2006c}.
They have also given us the first clues about the interior structures of
these other worlds. Most of the transiting extrasolar planets have
mean densities between 0.6 and 1.2~g~cm$^{-3}$, suggesting they are
not too different from the well-studied gas giants Saturn
(0.7~g~cm$^{-3}$) and Jupiter (1.3~g~cm$^{-3}$). However, the first
transiting planet that was discovered, HD~209458b, has a much smaller
density of 0.33~g~cm$^{-3}$~(\citealt{Henry.2000,Charbonneau.2000};
for the most recent analyses see
\citealt{Winn.2005b},\citealt{Wittenmyer.2005}, and
\citealt{Knutson.2006}). This anomaly has led to speculation about 
novel sources of internal heat, such as eccentricity damping 
\citep{Bodenheimer.2001}, insolation-driven weather patterns
\citep{Guillot.2002}, and obliquity tides \citep{Winn.2005a}. 

The periodic dimming events of OGLE-TR-10 were discovered by
\citet{Udalski.2002a} in a survey for transiting planets in three star
fields toward the Galactic center. \citet{Udalski.2002c} reported
additional observations that enabled the events to be predicted with
greater precision. Spectroscopic follow-up observations by
\citet{Konacki.2005}, in combination with those reported by
\citet{Konacki.2003b} and \citet{Bouchy.2005a}, revealed a periodic
Doppler shift and thereby confirmed that the dimming events were
caused by the transits of a Jovian planet.  The estimated planetary
mass is \mp.  Because of the planet's anomalously small mean density
of $0.38\pm 0.10$~g~cm$^{-3}$, the discovery of this system was
greeted with considerable interest
\citep{Konacki.2005,Bouchy.2005a,Baraffe.2005,Laughlin.2005,Gaudi.2005,Santos.2006}.

Through the recently initiated Transit Light Curve (TLC) Project, we
are building a library of high-precision transit photometry, with the
dual goals of (1) refining the estimates of the physical and orbital
parameters of the target systems, and (2) searching for secular and
short-term variations in the transit times and light curve shapes that
would be indicative of perturbations from additional
bodies~\citep{Agol.2005,Holman.2005a}.  We have previously reported on
observations of the exoplanets XO-1b~\citep{Holman.2006},
TrES-1~\citep{Winn.2006a}, and OGLE-TR-111b~\citep{Winn.2006b}.  In
this paper\footnote{This paper is a thorough revision of (and
supersedes) a previous version of this manuscript that was circulated
on astro-ph in 2005. The most important change is that we obtained
observations of 2 additional full transits to check our results.
The new data are in agreement with our original data, and our
conclusion is unchanged that the planetary radius is smaller than was
previously believed. See \S~5 for a discussion.}, we present TLC
results for OGLE-TR-10b. We describe our observations in \S~2 and data
reduction procedures in \S~3. In \S~4, we describe the modeling
procedures by which we estimate the physical and orbital parameters of
OGLE-TR-10. We present the results in \S~5 and summarize our findings
in \S~6.

\section{Observations and Data Reduction}

We used the 6.5m Baade (Magellan~I) and Clay (Magellan~II) telescopes
at Las Campanas Observatory, in northern Chile, to observe five
different transits of OGLE-TR-10.  These correspond to epochs $E=223$,
251, 252, 488, and 597 of the ephemeris determined by
\citet{Udalski.2002b}:
\begin{equation} T_c(E) =
2,452,070.21900~{\mathrm{[HJD]}} + E\times(3.10140~{\mathrm{days}}).
\end{equation}
We observed the first four transits with the Raymond and Beverly
Sackler Magellan Instant Camera (MagIC) on the Clay telescope.  MagIC
is a 2k~$\times$~2k SITe back-illuminated and thinned CCD with
24~$\mu$m pixels.  For transits $E=223$, 251, and 252, we alternated
between observations in Johnson-Cousins $B$ and $I$ filters. For
transit $E=488$, we used the $I$ filter only.  We used a typical
exposure time of 30-60~s. The readout and reset time was about 25~s
and the read noise was about 5~e$^{-}$~pixel$^{-1}$. The seeing
varied from 0.4--2\arcsec.

A low-amplitude herringbone pattern was evident in the MagIC
images. The same pattern appeared in each of the 4 quadrants of an
individual exposure, but the pattern varied randomly between
images. This is a known problem resulting from 60~Hz noise. For each
image, we created a herringbone template by taking the median of the
four quadrants after masking the stars and applying a narrow-band
60~Hz Fourier filter, based on an algorithm kindly provided by
S.~Burles. We then subtracted the isolated and purified herringbone
pattern from the images. A shutter correction was applied, and bad
pixels were assigned values based on linear interpolation of
neighboring pixel values.

We observed the fifth transit ($E=597$) with the Inamori Magellan
Areal Camera and Spectograph (IMACS) on the Baade telescope.  The
IMACS detector is a mosaic of eight 2k~$\times$~4k SITe
back-illuminated and thinned CCDs with 15~$\mu$m pixels.  We used the
``long'' ($f/4.3$) camera, giving a pixel scale of $0\farcs111$ and a
field of view of $15\farcm4$.  To reduce the readout time, we read
only one third of each chip, corresponding to the central $15\farcm4
\times 5\farcm1$ of the mosaic. The readout time was approximately
45~s and the readout noise was about 5~e$^{-}$. We observed through
the CTIO $I$~band filter.

All of the images, from both MagIC and IMACS, were overscan corrected,
trimmed, and divided by a flat-field image with standard
IRAF\footnote{IRAF is distributed by the National Optical Astronomy 
Observatories, which are operated by the Association of Universities
for Research in Astronomy, Inc., under cooperative agreement with the
National Science Foundation.} routines.  A bias frame was subtracted
from the IMACS images before the flat-field correction was applied
(this step was unnecessary for the MagIC images).  We obtained the dome
flat exposures and zero-second (bias) exposures at the beginning of
each night.

Because the region surrounding OGLE-TR-10 was too crowded for aperture photometry,
we used the method of image subtraction as implemented by
\citet{Alard.1998} and \citet{Alard.2000}. Specifically, we used version 2.2 of
the ISIS image subtraction package that was written and kindly made
public by C.~Alard. In this method, all of the images from a given
night are registered to a common pixel frame, and a reference image is
created by combining a subset ($\approx$10\%) of the images with the
best seeing from that night. For each individual image, a convolution kernel is
determined that brings the image into best agreement with the
reference image. Then the difference between the appropriately
convolved image, and the reference image, is computed. The advantage
of this method is that photometry is simplified on the difference
images, because most stars are not variable stars and thus the complex
and crowded background is eliminated. It is still necessary to compute
the flux of the variable stars on the reference image, taking into
account any neighboring stars, but this need only be done once, and
the task is facilitated by the good spatial resolution and high
signal-to-noise ratio of the reference image. Thus, the measurement of
the relative flux $f(t)$ takes the form
\begin{equation}
f(t) = \left[f_{\rm ref} + \Delta f(t)\right]/f_{\rm oot},
\end{equation}
where $f_{\rm ref}$ is measured on the reference image, $\Delta f(t)$
is measured on the difference images, and $f_{\rm oot}$ is the out of
transit flux.  We derived $f_{\rm ref}$ for each of the reference
images by using DAOPHOT~\citep{Stetson.1987} to measure the flux
through a small aperture with PSF fitting, and then calculating the
appropriate aperture correction by examining an ensemble of bright,
relatively isolated stars for which we had subtracted any remaining
nearby stars. 

Although the image subtraction method removes the first-order effects
of extinction by scaling all of the images to a common flux level
before subtraction, residual color-dependent effects are not removed.
Stars of different colors are extinguished by different amounts
through a given airmass $z$.  Thus, as part of the fitting procedure
discussed in \S~3, we apply a residual extinction correction to the
data such that the observed flux is proportional to $\exp(-kz)$, where
$k$ is a residual extinction coefficient.

The uncertainty in the each data point arises from two sources: the
uncertainty in the difference flux $\Delta f$, and the uncertainty in
the reference flux $f_{\rm ref}$. We estimated the uncertainty in the
difference flux based on Poisson statistics. This turned out to be an
underestimate; some correlated errors are evident in the final light
curves (see the next section).  We estimated the uncertainty in the
reference flux based not only on Poisson statistics, but also by the
spread in the values obtained when using different choices for the
stars used to determine the point-spread-function (PSF) and other
parameters relating to the profile photometry. This latter source of
systematic error was typically 1\%, which dominated the Poisson error
in the reference flux determination. However, adjustments in $f_{\rm
ref}$ affect all of the points from a given night in the same way; the
net effect is a small modification of the transit depth. For example,
for OGLE-TR-10 the transit depth is approximately 1\%. Because $f_{\rm
  oot} \sim f_{\rm ref}$, the effect of increasing $f_{\rm ref}$ by 1\% is to
decrease the transit depth by $(0.01 \times 0.01)$ or $10^{-4}$. As we
will show in the next section, this error proved to be smaller than
the other errors in the difference fluxes. In addition, there may be a
systematic error of order $3$~seconds ($3\times 10^{-5}$~days) in the
reported observation times, due to a delay between the opening of the
shutter and the recording of the time.  We have not verified or
attempted to correct for this delay.

\section{The Model}

Of the five nights of observations, the data from the two most recent
events ($E=488$ and 597) have the highest sampling frequency and most
uniform coverage and quality.  We determined the system parameters of
OGLE-TR-10b by fitting a model to these data only.  Our model and
fitting algorithm are similar to those employed in previous TLC
papers~\citep{Holman.2006, Winn.2006b, Winn.2006a}. The model posits a
circular Keplerian orbit of a star with mass $M_{\rm S}$ and radius
$R_{\rm S}$, and a planet with mass $M_{\rm P}$ and radius $R_{\rm P}$
with period $P$ and inclination $i$ relative to the sky
plane\footnote{Unless there is positive evidence for a nonzero
eccentricity, a circular orbit is a reasonable simplifying assumption
for a ``hot Jupiter'' around a main-sequence star.  This is because
(in the absence of any third body) it is expected that there has been
sufficient time for tides to have damped out any initial eccentricity,
in the absence of a third body (see, e.g.,
\citealt{Rasio.1996,Trilling.2000,Dobbs-Dixon.2004}.}. We define the
coordinate system such that $0\arcdeg \leq i\leq 90\arcdeg$.

The stellar mass cannot be determined from transit photometry alone.
There is a well-known degeneracy among $M_{\rm S}$, $R_{\rm S}$ and
$R_{\rm P}$ that prevents all three parameters from being uniquely
determined from transit photometry alone, unless a stellar mass-radius
relation is assumed~\citep{Seager.2003}. Our approach was to
fix \ms based on an analysis of the stellar spectrum (see the
discussion in \S~4). We then use the scaling relations $R_P \propto
M_S^{1/3}$ and $R_S \propto M_S^{1/3}$ to estimate the systematic
error due to the uncertainty in $M_S$. The planetary mass $M_{\rm P}$
is irrelevant to the model except for its minuscule effect on the
relation between $P$ and the semimajor axis; for completeness, we
assume \mp, the value reported by \citet{Konacki.2005}.

We allow each of the two fitted transits to have an independent value
of $T_c$, the transit midpoint, rather than forcing 
them to be separated by an integral number of orbital periods. Thus,
the period $P$ is relevant to the model only through the connection
between the total mass and the orbital semimajor axis. We fix
$P=3.10140$~days~\citep{Udalski.2002c}. 

The model flux is unity outside of transit, and is otherwise computed
using a quadratic limb darkening law, $I = 1 - u_1(1-\mu) -
u_2(1-\mu)^2$.  We employed the analytic formulas of
\citet{Mandel.2002} to compute the integral of the intensity over the
exposed portion of the stellar disk.  We allow the photometry of each
night (and in each filter) to have an independent out of transit flux
$f_{\rm oot}$ and residual extinction coefficient $k$.  We hold the
limb darkening parameters fixed at $u_1=0.2541$, $u_2=0.3254$ in $I$,
as appropriate for a star with the assumed properties\footnote{There
is a systematic error associated with the uncertainty in the limb
darkening function, but it is probably much smaller than the other
sources of error. For example, if we adopt the coefficients that are
appropriate to the hotter star favored by Bouchy et al.~(2005) and
Santos et al.~(2006), our inferred values of $R_P$ and $R_S$ change by
about 1.5\%, which can be neglected in comparison to the 6\%
statistical errors.} and a microturbulent velocity of
$2~\kms$~\citep{Claret.2000}.

In total, there are 9 adjustable parameters describing $N_f=601$
photometric data points. The parameters are $R_S$, $R_P$, and $i$; the
two values of $T_c$; and the values of $f_{\rm oot}$ and $k$ for each
transit. The goodness-of-fit parameter is
\begin{equation}
\chi^2 = \sum_{j=1}^{N_f}
\left[
\frac{f_j({\mathrm{obs}}) - f_j({\mathrm{calc}})}{\sigma_j}
\right]^2
\end{equation}
where $f_j$(obs) is the flux observed at time $j$, $\sigma_j$ is the
corresponding uncertainty, and $f_j$(calc) is the calculated value.

First we use an AMOEBA algorithm~\citep{Press.1992} to identify the
parameter values that minimize $\chi^2$.  We then rescale the
uncertainties $\sigma_j$ by a factor that is specific to each of the
two time series, such that $\chi^2/N_{\rm DOF} = 1$ for each time
series individually.  Then we estimate the {\it a posteriori} joint
probability distribution for the parameter values using a Markov Chain
Monte Carlo (MCMC) technique (see, e.g., \citealt{Tegmark.2004}). In
this method, a chain of points in parameter space is generated from an
initial point by iterating a jump function, which in our case was the
addition of a Gaussian random number to each parameter value. If the
new point has a lower $\chi^2$ than the previous point, the jump is
executed; if not, the jump is only executed with probability
$\exp(-\Delta\chi^2/2)$. We set the typical sizes of the random
perturbations such that $\sim$20\% of jumps are executed. We created
10 independent chains, each with 500,000 points, starting from random
initial positions. The first 100,000 points were not used, to minimize
the effect of the initial condition. The \citet{Gelman.1992} $R$
statistic was within 0.5\% of unity for each parameter, a sign of good
mixing and convergence.

Next we verify that our $BI$ photometry of events $E=223$, 251, and
252 (for which the time coverage was sparser) is consistent with the
model parameters derived above.  We fix the stellar and planetary
radii, inclination, and period at the best fit values. We hold the
limb-darkening coefficients fixed as before (with $u_1=0.6385$,
$u_2=0.1789$ in $B$).  We vary only the values of $f_{\rm oot}$ and
$k$ for each of the time series, as well as the three independent
values of $T_c$. The final photometry is given in Table~1, and is
plotted in Figs.~\ref{fig:lc1} and~\ref{fig:lc2}. For comparison, the
OGLE light curve is shown in Fig.~\ref{fig:ogle}.  The uncertainties
given in Table~1 are the uncertainties in the difference fluxes, after
multiplying by a factor specific to each night such that
$\chi^2/N_{\rm DOF} = 1$ for the best-fitting model.  The data from
individual transits are all generally consistent with one another.
Systematic errors are evident, especially in the $E=251$ data, but by
fitting for the residual extinction correction we have taken into
account the main systematic error that would afflict the determination
of $T_c$.

\section{The Results}

\begin{figure}[p]
\epsscale{0.8}
\plotone{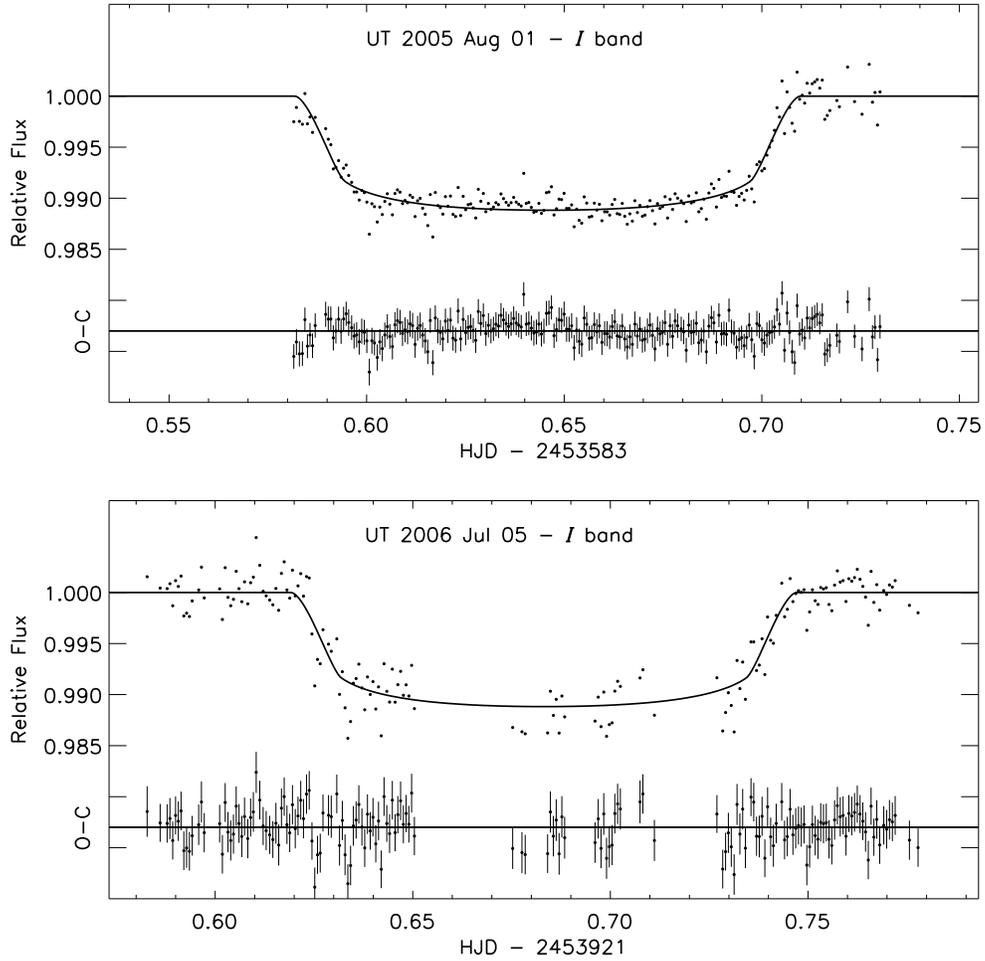}
\caption{Relative $I$ photometry of OGLE-TR-10 during the
$E=488$ and 597 transits.
The panels show the
photometry (points) and the best-fitting model (solid line), along
with the residuals (observed~$-$~calculated). These two data
sets were the ones used to determine the OGLE-TR-10 system
parameters because they had the best time sampling and accuracy.
\label{fig:lc1}}
\end{figure}

\begin{figure}[p]
\epsscale{0.8}
\plotone{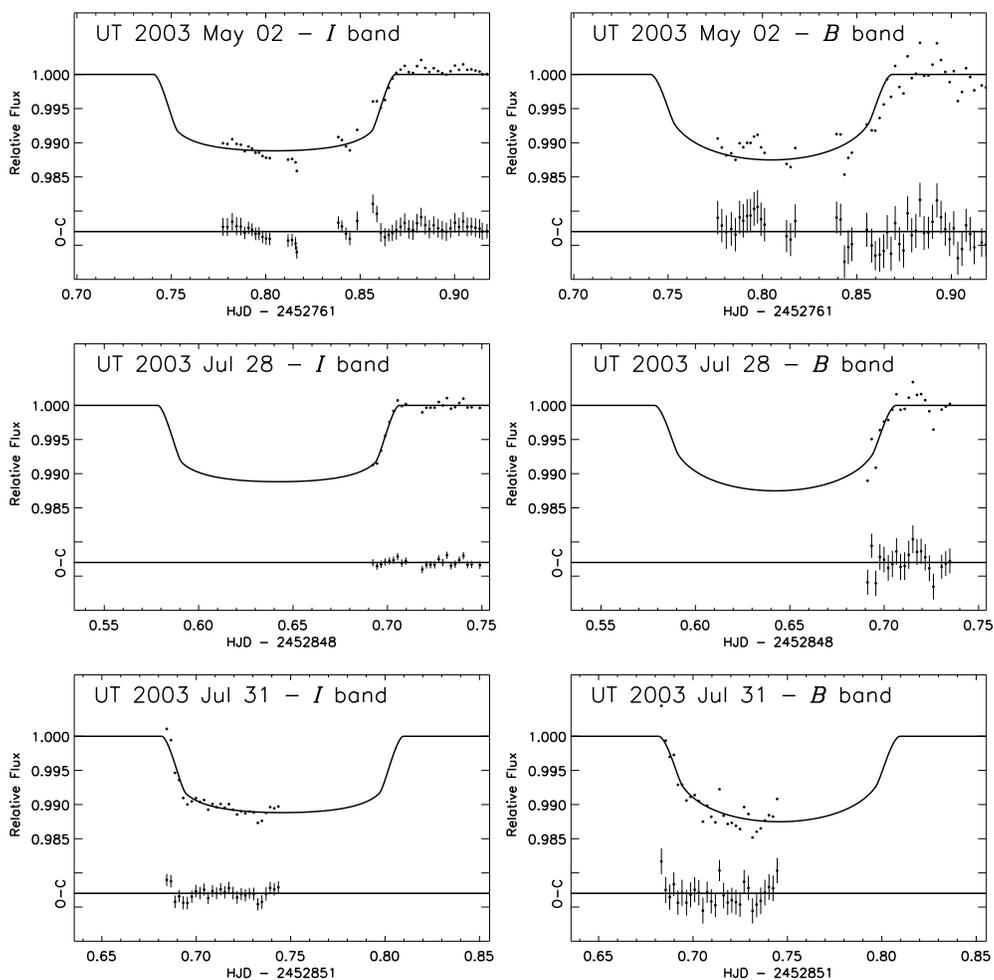}
\caption{Relative $BI$ photometry of OGLE-TR-10 during the
$E=223$, 251, and 252 transits.
The panels show the
photometry (points) and the best-fitting model (solid line), along
with the residuals (observed~$-$~calculated). These data
sets had incomplete coverage of the transits and were used
only as a consistency check and to estimate transit midpoint times.
\label{fig:lc2}}
\end{figure}

\begin{figure}[p]
\epsscale{0.8}
\plotone{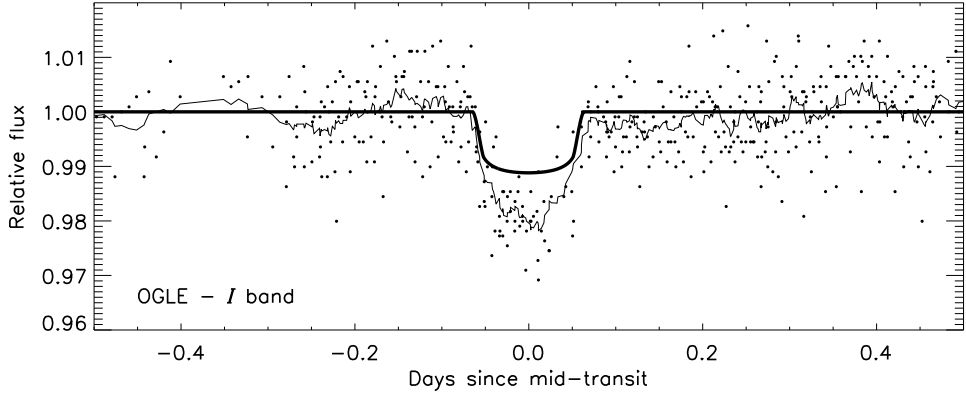}
\caption{Relative $I$ photometry of OGLE-TR-10, from the OGLE
discovery and follow-up data (Udalski et al.~2002ac).
The photometry (points) and our best-fitting model (solid line)
are plotted, showing that our model has a shallower
transit. The jagged line is a boxcar-smoothed version
of the data (with a boxcar size of 10 points), giving a visual
cue of the 0.5\% correlated errors in the OGLE photometry.
\label{fig:ogle}}
\end{figure}

\begin{figure}[p]
\epsscale{0.8}
\plotone{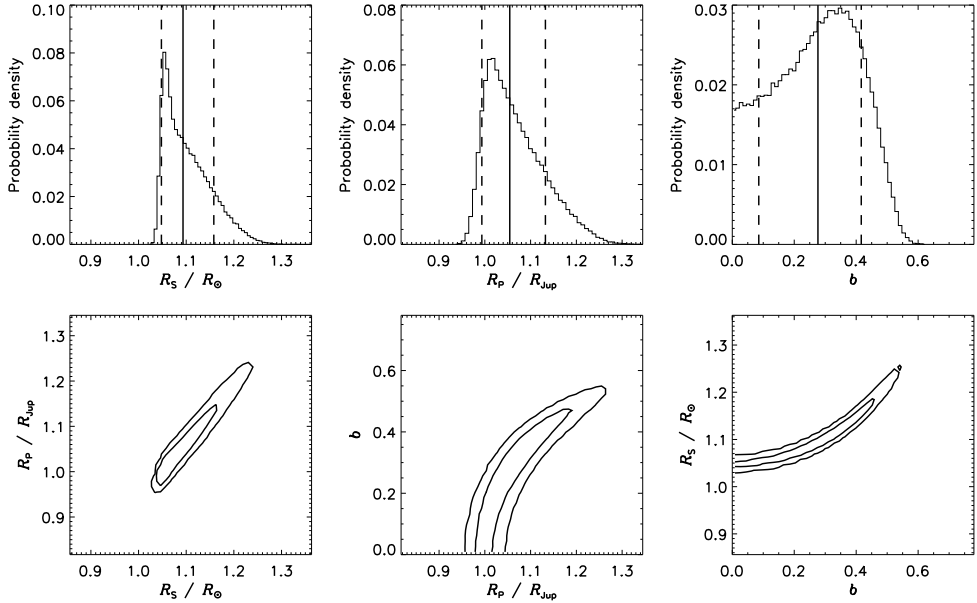}
\caption{ {\bf Top row.} Probability distributions for the stellar
  radius $R_S$, planetary radius $R_P$, and impact parameter $b\equiv
  a\cos i/R_S$, based on the MCMC simulations. The solid lines mark the
  median values; the dashed lines mark the 68\%  confidence limits.  
  {\bf Bottom row.}  Joint probability
  distributions of those parameters with the strongest correlations.
  The contours are isoprobability contours enclosing 68\% and 95\% of
  the points in the Markov chains. 
\label{fig:probdist}}
\end{figure}

The model that minimizes $\chi^2$ is plotted as a solid line in
Figs.~1-3. The optimized residual extinction correction has been
applied to the data that are plotted in Fig.~\ref{fig:lc1}, and to the
data that are given in Table~1. The differences between the observed
fluxes and the model fluxes are also shown beneath each light curve.
Table~2 gives the estimated values and uncertainties for each
parameter based on the MCMC analysis. It also includes some useful
derived quantities: the impact parameter $b= a \cos i / R_S$ (where
$a$ is the semimajor axis); the time between first and last contact
($t_{\rm IV} - t_{\rm I}$); and the time between first and second
contact ($t_{\rm II} - t_{\rm I}$). Fig.~\ref{fig:probdist} shows the
estimated {\it a posteriori} probability distributions for the
especially interesting parameters $R_S$, $R_P$ and $b$, along with
some of the two-dimensional correlations involving those
parameters. Although the distributions shown in
Fig.~\ref{fig:probdist} are somewhat asymmetric about the median,
Table~2 reports only the median $p_{\rm med}$ and a single number
$\sigma_p$ characterizing the width of the distribution. The value of
$\sigma_p$ is the average of $|p_{\rm med}-p_{\rm hi}|$ and $|p_{\rm
med} - p_{\rm lo}|$, where $p_{\rm lo}$ and $p_{\rm hi}$ are the lower
and upper 68\% confidence limits.

Our estimate of the planetary radius, \rp, is smaller than the
previous estimates of $R_{\rm P}/R_{\rm Jup} = 1.24\pm
0.09$~\citep{Konacki.2005} and $1.54\pm 0.12$~\citep{Bouchy.2004},
both of which were based on the OGLE discovery photometry.  It is
important to note that both of these earlier works used a value for
$R_S$ that was determined from an analysis of the stellar spectrum,
whereas we have determined $R_S$ by fitting the transit light curves.
One reason that the previous estimates of the planetary radius were
larger is the apparently larger value of the flux decrement in the
OGLE photometry (see Fig.~\ref{fig:lc1}).  The larger discrepancy with
\citet{Bouchy.2004} is also due to their assumption of a larger value
for the stellar radius.

We do not know why the OGLE photometry shows a larger decrement.  We
first saw this discrepancy based on our 2003 data (which was the basis
of an earlier version of this manuscript that was circulated on
astro-ph).  We could not find any errors in our photometry, but
because none of our light curves adequately covered a complete transit
we decided to obtain additional photometry in 2005 and 2006 to try to
resolve the matter. The transit depth in the newer and more complete
light curves is also shallower than in the OGLE light curve, and is in
agreement with our earlier data, despite differences in the telescope,
detector and observing conditions. Although there are systematic
errors in both our photometry and the OGLE photometry (see Figs.~1 and
2), one would expect our data to be more trustworthy, since they are
of higher precision and more frequent sampling. We also note that of
the seven transits of OGLE-TR-10 observed by OGLE, only two
(JD~2452085 and 2452113) were observed for a complete transit
including both pre-ingress and post-egress data.  Such data allow one
to visually check for systematic errors that might affect the measured
transit depth (which is why we obtained the two additional light
curves in 2005 and 2006).  The full transit on JD~2452085 is shallower
than the best model of \citet{Konacki.2005}, and the pre-ingress
observations on JD~2452113 are fainter than those post-egress.  The
OGLE team has consistently provided outstanding survey data, but the
data are necessarily reduced through a bulk photometry pipeline, as
opposed to the individual attention we have lavished on OGLE-TR-10.
Hence we proceed under the assumption that our photometry is correct.
A full resolution of this discrepancy would probably require
revisiting the original OGLE image frames.

Another point of disagreement regarding this system is the choice of
the stellar mass, which must be made independently of the transit
photometry.  Estimates of the stellar mass are based on analyses of
the stellar spectrum.  First, the values of $T_{\rm eff}$, $\log g$,
and metallicity are estimated by by measuring the equivalent widths
and profiles of appropriate lines and comparing them to
stellar-atmosphere models. Then, these derived parameters are
converted into a stellar mass, radius, and age, with reference to
theoretical models of stellar evolution (``isochrones'').

The two panels of figure~\ref{fig:isochrones} show Yonsei-Yale
isochrones~\citep{Yi.2003} for two respective metallicities of
OGLE-TR-10.  The left panel shows isochrones for the metal-rich value
measured by \citet{Santos.2006}; the right panel shows isochrones
assuming solar metallicity.  \citet{Santos.2006}, whose results
supercede those of \citet{Bouchy.2005a}, measure an effective
temperature $T_{\rm
  eff} = 6075 \pm 86$~K and a surface gravity $\log g = 4.54 \pm
0.15$.  These are indicated by the shaded rectangle.   The red line
shows the isochrone that best matches their results, without passing
below the ZAMS.  It is clear that this isochrone is only marginally consistent with
their estimates of the effective temperature and surface gravity.  In
addition, the best fitting isochrone corresponds to a very young stellar age of 0.2~Gyr.
Thus, we have re-analyzed the spectra of \citet{Konacki.2005}. We confirm
the measurement of solar metallicity ${\rm [Fe/H]}= 0.0\pm 0.2$. We also confirm
$\log g = 4.4^{+0.3}_{-0.7}$ and a slightly higher  $T_{\rm eff} = 5800 \pm
100$~K. We made wider use of temperature sensitive pairs of metal lines in deriving
the effective temperature (after \citealt{Gray.1991}). We used the following
lines: VI 625.183 and FeI 625.257 nm, SiI 612.503 and TiI 612.622 nm, 
VI 615.015 and FeI 615.162 nm. All three pairs gave consistent results.
In addition, we estimated $v\sin i = 3 \pm 1~{\rm km/sec}$, with microturbulent velocity
of $1.0~{\rm km/sec}$ and adopted (for this $T_{\rm eff}$) macroturbulent broadening of
$3.5~{\rm km/sec}$. In analogy to the planetary system HD~209458, we assume that
$v\sin i$ of a star with a transiting hot Jupiter is equal to the
rotational velocity $v_{rot}$. This allows
us to constrain the upper bound of the star's luminosity on the HR diagram,
by using an empirical relation between stellar age and rotation
\citep{Pace.2004}. The larger the rotational velocity of a G0~dwarf,
the younger and closer to the ZAMS it has to be. This constraint eliminates high luminosities
(and subgiant models) that are otherwise allowed for OGLE-TR-10, due
to the inability to constrain $\log g$ with sufficient accuracy with the
spectra. In addition, the luminosity can be constrained from below, as
models below the ZAMS of ${\rm [Fe/H]}= -0.2$ are excluded 
as well. These constraints, converted back to $\log g$ in order to aid the
illustration, are displayed in the right panel of Fig.~\ref{fig:isochrones}.
The red line in the right panel shows the isochrone that best fits the
constraints on $T_{\rm eff}$, $\log g$, and  ${\rm [Fe/H]}$.  This
isochrone corresponds to a 5.0~Gyr stellar age.  This consistency
gives us confidence in our measurements of effective temperature and
surface gravity.  We find the corresponding stellar mass to be $M_S =
1.025^{+0.125}_{-0.120}$~$M_\odot$.  To this should be added a 
  systematic error of $\sim 0.05~M_\odot$ due to the unknown helium 
abundance and convective mixing-length parameter. The theoretical
radius estimate is $R_{\rm S}/R_\odot = 1.057^{+0.194}_{-0.160}$.  

Fortunately, the fitted radius estimates depend weakly on the stellar
mass.  As explained in \S~3, our procedure was to assume $M_{\rm
S}/M_\odot = 1.025$ and allow the stellar radius to be a free
parameter. The result was \rs, which is consistent with the
theoretical value above.  If we assume the \citet{Santos.2006} value
of the stellar mass, $M_{\rm S}/M_\odot = 1.17$, we find $R_{\rm S} =
1.151 \pm 0.065$~$R_\odot$, which is also consistent with both the
theoretical value determined
by \citet{Santos.2006}, $R_{\rm S} = 1.144 \pm 0.062$~$R_\odot$ , as
well as with our value.  The larger value of the stellar
mass also has negligible effect on the estimate of the planetary radius:
$R_{\rm P} = 1.103 \pm 0.075$~$R_{\rm Jup}$ in comparison to the value
\rp ~determined with our mass estimate.  Although a more accurate
determination of the stellar mass would be necessary to take full
advantage of higher precision photometry, such as that obtained with
the HST, it is not needed for this work.

We fitted a straight line to the transit times ($T_c$) listed in Table~2
to derive a new ephemeris
\begin{equation}
T_c(E) = 2453921.684(1)~{\rm HJD} + E\times 3.101278(4)~{\rm days},
\end{equation}  
where the numbers in parentheses are the 1$\sigma$ uncertainties in the
final quoted digits.  Evaluating the dynamical significance of the
transit times, as was done for TrES-1 by \citet{Steffen.2005} and for
HD~209458b by \citet{Miller-Ricci.2005}, is left for a future investigation.

\section{Discussion}

Our results provide further evidence that the transits of OGLE-TR-10
are caused by a short-period planetary companion. The light curves
have a flat bottom, a feature that is inconsistent with most of the
alternative hypotheses involving blends of an eclipsing binary. As
further evidence against a blend, OGLE-TR-10 appears unresolved even
in the images with the best seeing (0\farcs 4).

It was previously thought that OGLE-TR-10b has a radius that is
significantly larger than predicted by models of strongly irradiated
gas giants. This would make it similar to the first-discovered
transiting planet HD~209458b, perhaps heralding a new subclass of
planets that share a mysterious internal heating mechanism.  The
suggestion that OGLE-TR-10 is anomalously large was made by
\citet{Konacki.2005} and was supported with calculations by
\citet{Baraffe.2005} and \citet{Laughlin.2005}, who found that the
measured radii of all the transiting planets could be easily explained
except for HD~209458b and OGLE-TR-10. Likewise, \citet{Gaudi.2005}
derived radii of $1.04~R_{\rm Jup}$ and $1.18~R_{\rm Jup}$ for
OGLE-TR-10 with and without a core, respectively, from the models
presented in \citet{Bodenheimer.2003}. These estimates include a 5\%
increase to account for the effect pointed out by
\citet{Burrows.2003}, that the measured radius of a transiting planet
refers to the scale height of the planetary atmosphere at which the
stellar flux encounters optical depth $\sim$1 in the direction toward
the observer. Our results indicate that OGLE-TR-10b does not have an
anomalously large radius, and that its radius is indeed consistent
with standard structural models.

\acknowledgments We are grateful to S.\ Burles for sharing his
algorithm for Fourier filtering of MagIC images. W.F.\ thanks the SAO
Summer Intern Program. M.J.H.\ and R.L.A.\ acknowledge support from
STScI grants GO~9433.05 and GO~9433.06. Work by J.N.W.\ was supported
by NASA through grant HST-HF-01180.02-A. Work by G.T.\ was supported
by NASA Origins grant NNG04GL89G.

\clearpage
\begin{deluxetable}{lcccc}
\tabletypesize{\normalsize}
\tablecaption{Photometry of OGLE-TR-10\label{tbl:photometry}}
\tablewidth{0pt}

\tablehead{
\colhead{Instrument} & \colhead{Filter} & \colhead{HJD} & \colhead{Relative flux} & \colhead{Uncertainty}
}

\startdata
     IMACS & $I$ &   2453921.58279 &         1.00140 &         0.00247 \\
\enddata 

\tablecomments{The time stamps represent the Heliocentric Julian Date
  at the time of mid-exposure. The uncertainty estimates are based on
  the procedures described in \S~2. We intend for this table to appear
  in entirety in the electronic version of the journal. A portion is
  shown here to illustrate its format. The data are also available
  from the authors upon request.}

\end{deluxetable}

\begin{deluxetable}{ccccc}
\tabletypesize{\normalsize}
\tablecaption{System parameters of OGLE-TR-10\label{tbl:params}}
\tablewidth{0pt}

\tablehead{
\colhead{Parameter} & \colhead{Value} & \colhead{Stat.\ Error} & \colhead{Sys.\ Error due to $M_S$} & \colhead{Total Error}
}

\startdata
                $R_S/R_\odot$& $          1.095$ & $          0.055$ &    $0.048$ &    $0.073$ \\
            $R_P/R_{\rm Jup}$& $          1.056$ & $          0.069$ &    $0.046$ &    $0.083$ \\
                  $R_P / R_S$& $         0.0990$ & $         0.0021$ &    \nodata &    \nodata \\
                    $i$~[deg]& $           88.1$ & $            1.2$ &    \nodata &    \nodata \\
                          $b$& $           0.28$ & $           0.17$ &    \nodata &    \nodata \\
$t_{\rm IV} - t_{\rm I}$~[hr]& $          3.072$ & $          0.050$ &    \nodata &    \nodata \\
$t_{\rm II} - t_{\rm I}$~[min]&$           17.9$ & $            2.0$ &    \nodata &    \nodata \\
             $T_c(597)$~[HJD]& $  2453921.68315$ & $        0.00056$ &    \nodata &    \nodata \\
             $T_c(488)$~[HJD]& $  2453583.64572$ & $        0.00053$ &    \nodata &    \nodata \\
             $T_c(252)$~[HJD]& $  2452851.74563$ & $        0.00064$ &    \nodata &    \nodata \\
             $T_c(251)$~[HJD]& $  2452848.64208$ & $        0.00043$ &    \nodata &    \nodata \\
             $T_c(223)$~[HJD]& $  2452761.80544$ & $        0.00091$ &    \nodata &    \nodata
\enddata

\tablecomments{
The parameter values in Column 2 are the median values $p_{\rm med}$
of the MCMC distributions. The statistical error given in Column 3 is
the average of $|p_{\rm med}-p_{\rm lo}|$ and $|p_{\rm med}-p_{\rm
hi}|$, where $p_{\rm lo}$ and $p_{\rm hi}$ are the lower and upper
68\% confidence limits.  These results are based upon the assumption
$M_{\rm S}/M_\odot = 1.025$ exactly.  The error given in Column 4 is
the additional systematic error due to the covariance with the stellar
mass, which was taken to be \ms.  Column 5 shows the quadrature sum of
the statistical and systematic errors for the stellar and planetary radii.
}

\end{deluxetable}

\begin{figure}[p]
\epsscale{1.0}
\plotone{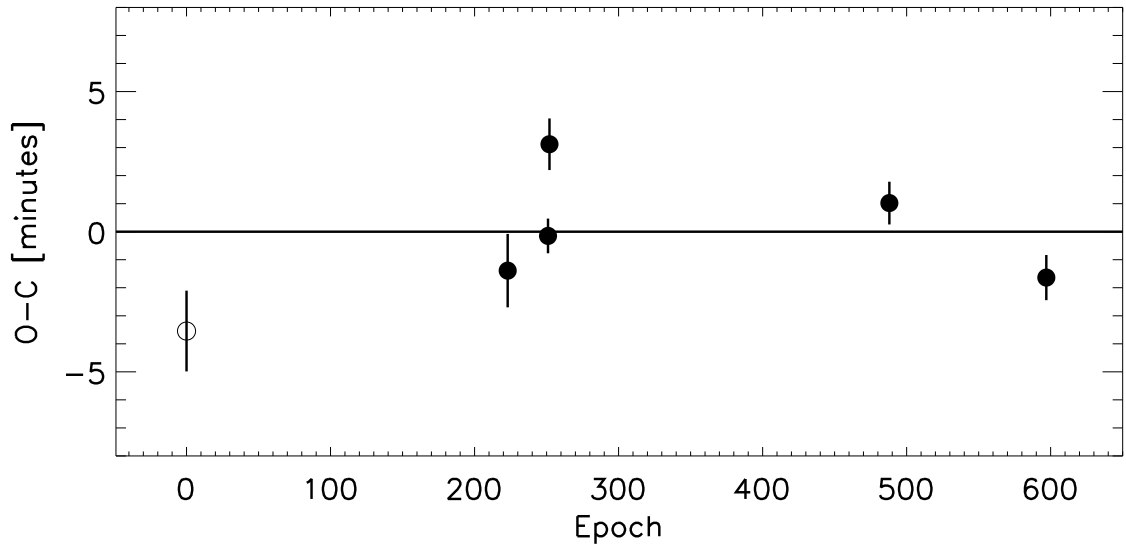}
\caption{Transit timing residuals for OGLE-TR-10. The
  calculated transit times have been subtracted from the observed
  transit times.  The open symbol indicates the $T_c$ from
  \citet{Udalski.2002c}; we estimate its uncertainty to be 0.001~days.
   The solid symbols indicate our measurements.
  The best-fitting line is plotted, representing the
  updated ephemeris given in Eq.~4 and the text that follows it. 
\label{fig:tc}}
\end{figure}

\begin{figure}[p]
\epsscale{1.0}
\plotone{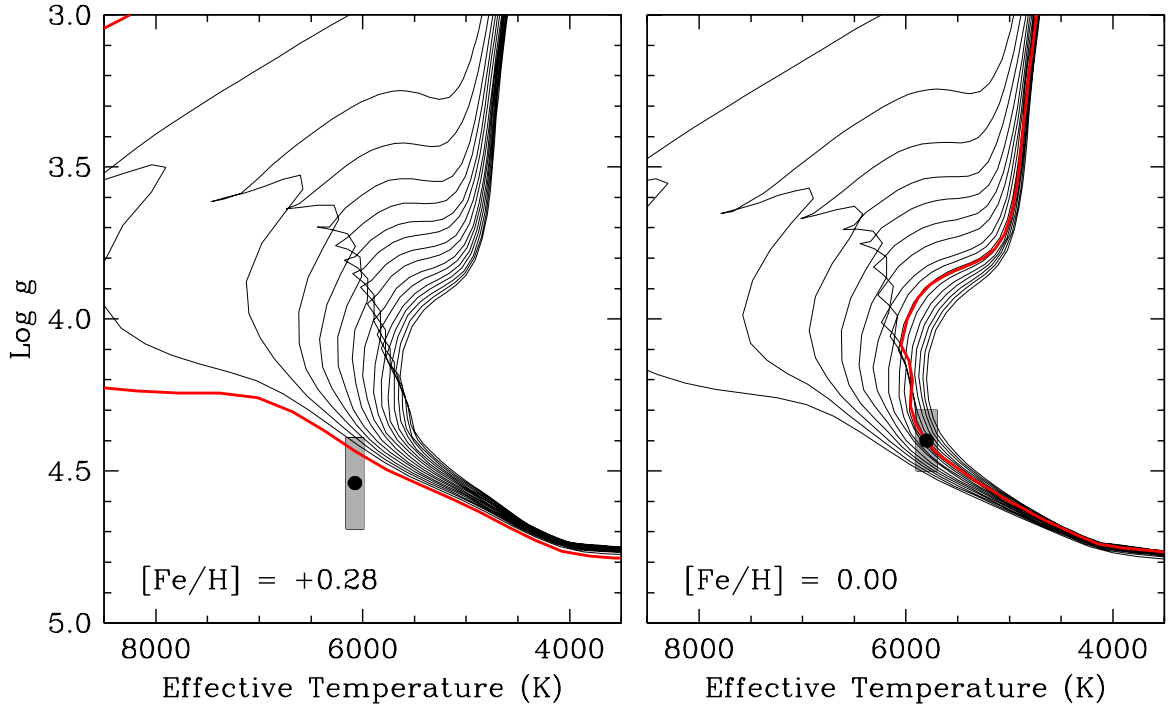}
\caption{ Yonsei-Yale isochrones as a function of effective
  temperature and surface gravity \citep{Yi.2003}.  {\bf Left.} Isochrones for the metallicity determined by
  \citet{Santos.2006}.  The black point marks the best-fit effective
  temperature and surface gravity measured in that work.  The shaded
  rectangle indicates the uncertainties in effective temperature and
  surface gravity. The red line shows the isochrone that best fits the constraints on $T_{\rm eff}$, $\log g$,
  and  ${\rm [Fe/H]}$, without passing   below the ZAMS.  It
  corresponds to  a stellar age of 0.2~Gyr and a   metal content ${\rm
  [Fe/H]}= 0.18$.   {\bf Right.} Corresponding isochrones for   solar
  metallicity, as determined by   \citet{Konacki.2005} and by 
  this work. The red line corresponds to a stellar age of 5.0~Gyr and
  a metallicity ${\rm [Fe/H]}= 0.04$.
\label{fig:isochrones}}
\end{figure}

\end{document}